\documentclass[conference]{IEEEtran}

\usepackage{cite}
\usepackage{amsmath,amssymb,amsfonts}
\usepackage{graphicx}
\usepackage{xcolor}
\usepackage{booktabs}
\usepackage{subcaption} 
\usepackage{placeins}
\usepackage{adjustbox}
\usepackage[final]{microtype}
\begin{document}
\bstctlcite{IEEEexample:BSTcontrol}
\makeatletter
\newcommand{\name}[1]{\gdef\@ieeename{#1}}
\newcommand{\address}[1]{\gdef\@ieeeaddr{#1}\author{\@ieeename\\\@ieeeaddr}}
\makeatother

\title{MemNMF: Memory-Augmented NMF on LPC Spectra for Anomalous Sound Detection}

\name{Phurich Saengthong, Takahiro Shinozaki}
\address{Institute of Science Tokyo, Japan}

\maketitle

\begin{abstract}
Autoencoder-based anomalous sound detection is attractive for machine condition monitoring because it can be trained using only normal recordings and yields an interpretable anomaly score from reconstruction error. Most prior work uses spectrogram autoencoders, but reconstructing detailed time--frequency patterns is sensitive to noise and transients, and models can reconstruct some anomalous inputs well, weakening normal--anomaly separation. We propose MemNMF, a constrained reconstruction method that operates on the Linear Predictive Coding spectrum, a compact estimate of the spectral envelope. MemNMF initializes a memory module from an NMF dictionary learned on normal LPC spectra and reconstructs each input as an attention-weighted combination of prototypical normal spectral patterns. Experiments on MIMII and DCASE 2020 Task 2 across multiple machine types and operating conditions show that LPC-spectrum inputs improve a standard autoencoder baseline and that MemNMF yields further gains, with especially strong robustness under noisy, non-stationary settings.
\end{abstract}

\begin{IEEEkeywords}
anomalous sound detection, autoencoder, linear predictive coding, non-negative matrix factorization, memory network.
\end{IEEEkeywords}

\section{Introduction}
Anomalous sound detection (ASD) aims to detect faults from machine sounds, enabling early maintenance and preventing costly downtime~\cite{Koizumi_8081297, Uematsu2017AnomalyDT, purohit_mimii_2019, koizumi_toyadmos_2019, koizumi_description_2020}. Since anomalous events are rare and diverse, practical systems are often trained only on normal recordings and must detect unseen failures at test time~\cite{koizumi_description_2020}. In this setting, reconstruction-based approaches such as autoencoders (AEs) are attractive because they are simple to develop, require only normal data, and provide an end-to-end anomaly score via reconstruction error~\cite{Koizumi_8081297, koizumi_description_2020}; moreover, reconstruction errors can provide an interpretable anomaly signal in the input space.

Despite their simplicity, AE-based approaches can struggle on noisy, non-stationary recordings, where the reconstruction objective may fit spurious variations rather than invariant machine characteristics. With time-varying spectrogram inputs, the model must reconstruct fine-grained time--frequency detail; when the observed spectrum varies substantially over time, this target diversity encourages fitting nuisance variation rather than the underlying structure that defines the normal machine state, making it difficult to learn a stable normal representation~\cite{purohit_mimii_2019, suefusa_anomalous_2020}. Moreover, because decoding from a continuous latent space is only weakly constrained, the model may also reconstruct patterns not supported by normal data, reducing normal--anomaly separability~\cite{gong_memorizing_2019}.

In this paper, we cast robust reconstruction as a constrained representation problem: the model should reconstruct only from patterns supported by normal data and should operate in a feature space that emphasizes invariant machine structure. To this end, we replace time-varying spectrogram textures with the Linear Predictive Coding (LPC) spectrum, which focuses on the spectral envelope, and propose Memory Network-based NMF (MemNMF). We learn an NMF dictionary from normal LPC spectra and use it to initialize a memory network, while learning lightweight key--value projections that map each query to a small set of prototypical normal patterns. The memory then reconstructs an input LPC spectrum as a weighted combination of these patterns, explicitly constraining reconstruction rather than relying on an unconstrained decoder. Experiments on widely used ASD benchmarks, including MIMII \cite{purohit_mimii_2019, koizumi_toyadmos_2019} and DCASE2020 Task~2 \cite{koizumi_description_2020}, show that replacing spectrogram inputs with the LPC spectrum improves a standard AE baseline under noisy, non-stationary machine sounds, and that combining the LPC spectrum with MemNMF yields further gains.

Our contributions are as follows: (i) We show that using the LPC spectrum as the input representation for AE-style ASD can substantially improve performance on non-stationary machine sounds, highlighting the benefit of focusing on the spectral envelope. (ii) We propose MemNMF, which initializes a memory network with an NMF dictionary learned from normal data and constrains reconstruction through attention over prototypical normal spectral patterns. (iii) We conduct an empirical evaluation on MIMII and DCASE2020 Task 2 across multiple machine types and operating conditions, demonstrating competitive performance and improved robustness, particularly under noisy and non-stationary settings.

\begin{figure*}[!t]
    \centering\includegraphics[width=0.85\textwidth]{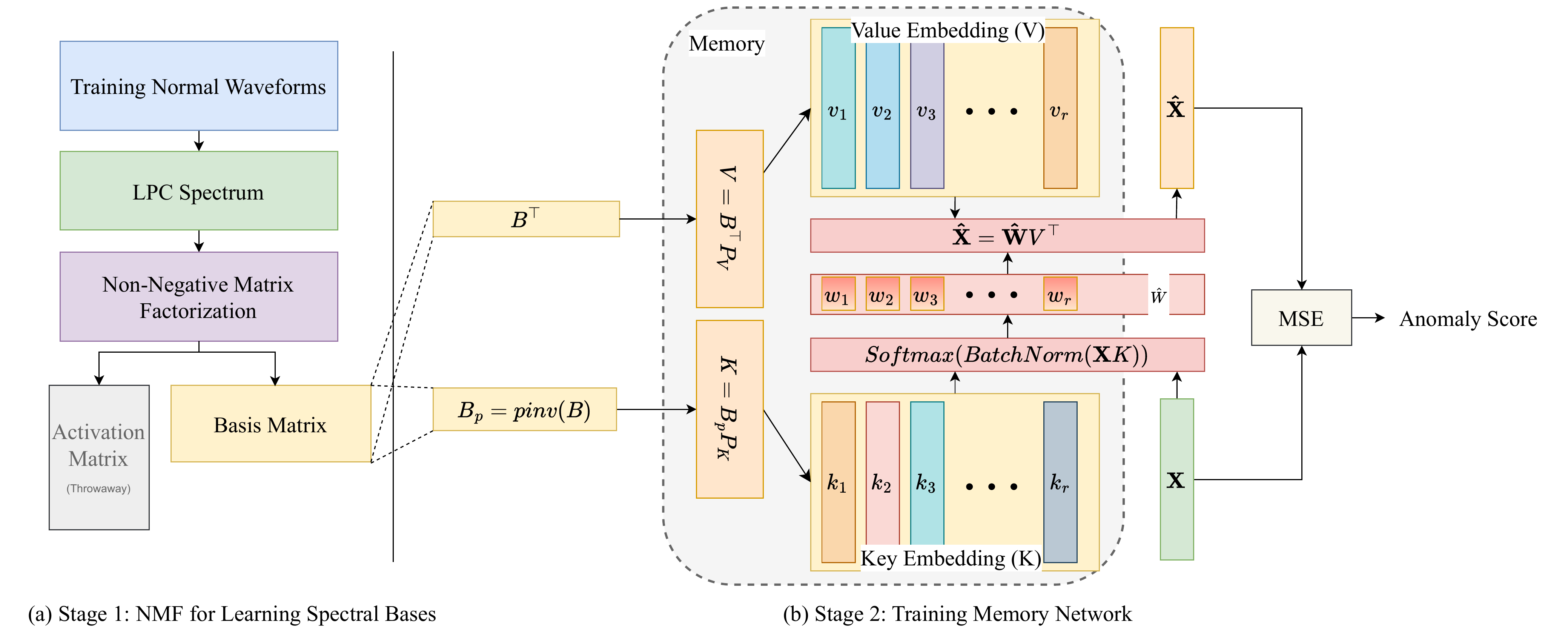}
    \caption{Overview of MemNMF: Stage 1 learns non-negative spectral bases from normal LPC spectra using NMF; Stage 2 trains projections for key--value embeddings derived from the basis vectors and a BatchNorm layer that normalizes dot-product logits to encourage sparse memory retrieval, using mean squared reconstruction loss.}
    \label{fig:memnmf}
    \vspace{-3mm}
\end{figure*}

\section{Related Work}
Most AE-based ASD systems operate on time--frequency representations, typically log-mel spectrograms \cite{Koizumi_8081297, koizumi_spidernet_9053620, purohit_mimii_2019, koizumi_description_2020, suefusa_anomalous_2020, wichern_anomalous_2021, zeng_joint_2023, zavrrtanki_10447941}. To improve AE-based ASD, several works modify the reconstruction objective to better capture temporal dependencies and reduce over-generalization. AE-IDNN interpolates centrally masked spectrogram frames from surrounding frames \cite{suefusa_anomalous_2020}; ANPs extend interpolation with flexible masking and attention over context frames \cite{wichern_anomalous_2021}; and PAE incorporates transformer layers while retaining an interpolation-style objective \cite{zeng_joint_2023}. Beyond spectrograms, \cite{GleichmannTNT2024} uses LPC-derived features and trains a VAE to reconstruct segment-level LPC. However, AE-based ASD that reconstructs an envelope-focused spectrum appears less commonly studied, even though the spectral envelope can provide a compact summary of spectral shape and potentially reduce sensitivity to frame-level fluctuations. From a broader reconstruction-based viewpoint, classical PCA-based anomaly detection also uses reconstruction error, but reconstructs from a low-dimensional orthogonal subspace \cite{huang_network_2006}; in contrast, MemNMF uses non-negative, NMF-initialized spectral bases on envelope-focused features. In addition, memory-augmented reconstruction has been studied for image anomaly detection \cite{gong_memorizing_2019}, but its use as an acoustic spectral-envelope dictionary for ASD remains less explored.

\FloatBarrier
\section{Method}
MemNMF addresses two limitations of spectrogram-based AEs by (i) using an LPC-spectrum representation that summarizes each clip by its spectral envelope, avoiding the need to reconstruct fine-grained, time--frequency textures, and (ii) constraining reconstruction to normal patterns via an NMF-derived memory learned from normal data so that atypical spectral structure yields larger reconstruction error. We learn NMF bases from normal LPC spectra and store them in a key--value memory. At test time, the memory reconstructs an input LPC spectrum as a weighted combination of stored items, and the reconstruction error serves as the anomaly score. Figure~\ref{fig:memnmf} illustrates the overall framework.

\subsection{Linear Predictive Coding (LPC) Spectrum}
LPC is a classic technique for modeling spectral envelopes \cite{saito1967theoretical, andersen1974calculation, burg1975maximum, lpc_review_1451722}. Importantly, we compute a single LPC model per audio clip and do not split the signal into short-time frames or apply windowing, as in spectrogram-based AEs. For each audio clip, we estimate LPC coefficients \(\{a_k\}_{k=1}^{p}\) directly from the time-domain waveform using Burg's method \cite{burg1975maximum, andersen1974calculation}. We then evaluate the corresponding all-pole model on \(m\) frequency bins \(\{\omega_i\}_{i=1}^{m}\) on the unit circle, yielding one LPC spectrum vector \(\mathbf{x} \in (\mathbb{R}_{\geq 0})^{m}\) per audio clip:
\begin{equation}
    x_i = |H(e^{j\omega_i})| = \left| \frac{1}{1 - \sum_{k=1}^{p} a_k e^{-j\omega_i k}} \right|, \quad i=1,\dots,m,
\end{equation}

\vspace{-2mm}
\subsection{Memory-Network-Based Non-Negative Matrix Factorization}
\subsubsection{Non-Negative Matrix Factorization}
\label{sec:nmf}

Given a matrix of normal LPC spectra \(\mathbf{X} \in (\mathbb{R}_{\geq 0})^{n \times m}\), where \(n\) is the number of training clips and \(m\) is the number of frequency bins, we learn a non-negative factorization~\cite{lee1999nmf}:
\begin{equation}
    \mathbf{X} \approx WB,
\end{equation}
where \(B \in (\mathbb{R}_{\geq 0})^{r \times m}\) contains \(r\) non-negative basis spectra that capture typical normal spectral patterns and \(W \in (\mathbb{R}_{\geq 0})^{n \times r}\) contains their activations. We optimize \(W\) and \(B\) under non-negativity using coordinate descent \cite{Cichocki2009FastLA} and then fix \(B\). We compute the Moore--Penrose pseudo-inverse \(B_p \in \mathbb{R}^{m \times r}\) as an analysis operator for memory-key initialization:
\begin{equation}
    B_p \;=\; \operatorname{pinv}(B).
\end{equation}

If \(B\) is full row-rank, a closed form is \(B_p = B^{T}(BB^{T})^{-1}\). Using \(B_p\), one can obtain approximate activations via \(\tilde{W}=\mathbf{X} B_p\) and reconstruct \(\tilde{\mathbf{X}}=\tilde{W}B\). For a single query \(\mathbf{x}\), the corresponding activation is \(\tilde{\mathbf{w}}=\mathbf{x} B_p\). From the analysis/synthesis viewpoint~\cite{mallat2009wavelet}, \(B\) can be regarded as a synthesis dictionary, while \(B_p\) acts as its dual analysis operator for computing activations. In the special case where \(BB^\top = I_r\), we have \(B_p = B^\top\). Unlike NNLS, the standard constrained NMF inference rule~\cite{lee2000algorithms}, \(\mathbf{x}B_p\) gives an unconstrained least-squares estimate, i.e., \(\tilde{\mathbf{w}}=\arg\min_{\mathbf{w}\in\mathbb{R}^{1\times r}}\|\mathbf{x}-\mathbf{w}B\|_2^2\), without enforcing \(\mathbf{w}\ge 0\). Thus, \(B_p\) is used not as a standalone inference method but as analysis-style key initialization, while MemNMF learns a normalized memory lookup anchored by \(B\).

\subsubsection{Memory Network Training}

We train the memory network on normal LPC spectra \(\mathbf{X} \in (\mathbb{R}_{\geq 0})^{n \times m}\) using a lookup over \(r\) memory items anchored to the fixed NMF dictionary \(B\). The key side uses the analysis operator \(B_p\), and the value side uses the synthesis dictionary \(B\): \(K = B_p P_K \in \mathbb{R}^{m \times r}\) and \(V = B^{\top} P_V \in \mathbb{R}^{m \times r}\), where \(P_K, P_V \in \mathbb{R}^{r \times r}\) are learnable projections. During training, we optimize \(P_K\), \(P_V\), and the BatchNorm parameters (including running statistics), while keeping \(B\) and \(B_p\) fixed.

To compute activations over memory items, we apply Softmax to the dot-product logits \(\mathbf{X}K\). During training, we apply BatchNorm to \(\mathbf{X}K\) before Softmax to normalize logit scale across samples. Accordingly, we compute
\begin{equation}
    \hat{\mathbf{W}} = \text{Softmax}(\text{BatchNorm}(\mathbf{X}K)),
\end{equation}
where Softmax is applied along the memory dimension to yield attention weights over the \(r\) memory items.

The corresponding reconstruction is \(\hat{\mathbf{X}} \in \mathbb{R}^{n \times m}\),
\begin{equation}
    \hat{\mathbf{X}} = \hat{\mathbf{W}}V^{\top}.
\end{equation}

We train on normal data by minimizing the mean squared reconstruction loss:
\begin{equation}
    \mathcal{L}_{\text{rec}} = \frac{1}{m}\|\mathbf{X} - \hat{\mathbf{X}}\|_F^2.
\end{equation}

At inference, we use \(\mathcal{L}_{\text{rec}}\) as the anomaly score.

\section{Experiments}
\begin{table}[t]
    \centering
    \caption{Comparison between AE and the proposed method on the sec-dev of DCASE2020T2 with AUC scores.}
    \label{tab:dcase2020_mimii_subset}
    \scriptsize
    \resizebox{\columnwidth}{!}{%
        \begin{tabular}{lccc}
            \toprule
            Machine & AE + Log-Mel & AE + LPC & MemNMF + LPC \\
            \midrule
            Toy-car & 80.9 & 71.3 & \textbf{86.2} \\
            Toy-conveyor & \textbf{73.4} & 55.4 & 62.3 \\
            Fan & 66.2 & 61.6 & \textbf{71.7} \\
            Pump & 72.9 & 75.4 & \textbf{78.8} \\
            Slider & 85.5 & 93.4 & \textbf{93.7} \\
            Valve & 66.3 & 96.8 & \textbf{97.6} \\
            \midrule
            Avg. AUC & 74.2 & 75.6 & \textbf{81.7} \\
            \bottomrule
        \end{tabular}%
    }
    \vspace{-4mm}
\end{table}

\subsection{Setup}

\noindent\textbf{Datasets.}
We evaluated on DCASE2020 Task~2 \cite{koizumi_description_2020}, which combines real-machine (MIMII subset \cite{Purohit_DCASE2019_01}) and toy-machine (ToyADMOS subset \cite{koizumi_toyadmos_2019}) recordings, all recorded with real factory noise across six machine types, and provides development and evaluation splits (\texttt{sec-dev}/\texttt{sec-eval}). We also reported results on the full MIMII dataset \cite{Purohit_DCASE2019_01}, which includes the overlapping real-machine types and three SNR conditions ($-6$, 0, and 6~dB). Following \cite{regional_local}, we used 11{,}419 segments for training and 6{,}600 for testing. All recordings are single-channel audio sampled at 16~kHz and are approximately 10~s long.

\noindent\textbf{Implementation details.}
We extracted 60 LPC coefficients using librosa \cite{brian_mcfee-proc-scipy-2015} and computed the LPC spectrum as an 8000-dimensional input feature. For NMF, we used 768 components. The memory network was trained with Adam \cite{Kingma2014AdamAM} and cosine learning-rate scheduling \cite{loshchilov2017sgdr} for 500 epochs, with batch size 200 and an initial learning rate of 0.001. For the LPC-spectrum AE baseline, we followed the AE implementation in \cite{Harada_EUSIPCO2023_01}, which originally used spectrogram inputs, and replaced the input and output layers to accept the 8000-dimensional LPC spectrum; we used batch size 512 and learning rate 0.01.

\noindent\textbf{Metrics.}
We reported the area under the receiver operating characteristic curve (AUC) and the partial AUC (pAUC) over a low false positive rate range \([0,p]\), following \cite{koizumi_description_2020}. In particular, pAUC computes the same ranking-based quantity but restricts evaluation to \(\mathrm{FPR} \le p\), emphasizing performance under strict false-alarm constraints. We used \(p=0.1\) \cite{koizumi_description_2020} and reported the arithmetic mean of AUC and pAUC across all machine sections \cite{local_den_norm_kevin}. Following \cite{Purohit_DCASE2019_01}, we trained and evaluated per section for each machine type.

\begin{table}[!t]
\centering
\caption{Comparison of AE-based systems on DCASE2020T2 using Amean over AUC and pAUC, reported for \texttt{sec-dev} and \texttt{sec-eval}\@.}
\label{tab:results_all}
\begin{adjustbox}{width=0.80\columnwidth}
\begin{tabular}{l@{\hskip 0.3cm}lll@{}}
\toprule
 Model & sec-dev & sec-eval & Amean \\
\midrule
AE \cite{koizumi_description_2020} &66.6 &70.0 &68.3 \\
AE-IDNN \cite{suefusa_anomalous_2020} &71.5 &- &- \\
ANP-IDNN \cite{wichern_anomalous_2021} & 71.8 & \textbf{75.6} & \textbf{73.7} \\
PAE \cite{zeng_joint_2023} &74.2 &- &- \\
AudDSR \cite{zavrrtanki_10447941} & \textbf{78.2} &- &- \\
\midrule
Ours (w/ LPC Spectrum) \\
AE & 73.6 & 72.8 &73.2 \\
MemNMF & \textbf{74.9} & \textbf{75.6} & \textbf{75.3} \\
\bottomrule
\end{tabular}
\end{adjustbox}
\vspace{-4mm}
\end{table}
\begin{figure*}[!t]
    \centering
    \setlength{\tabcolsep}{2pt}
    \begin{tabular}{@{}ccc@{}}
        \begin{subfigure}[t]{0.325\textwidth}
            \centering
            \includegraphics[width=\linewidth,height=4.6cm,keepaspectratio]{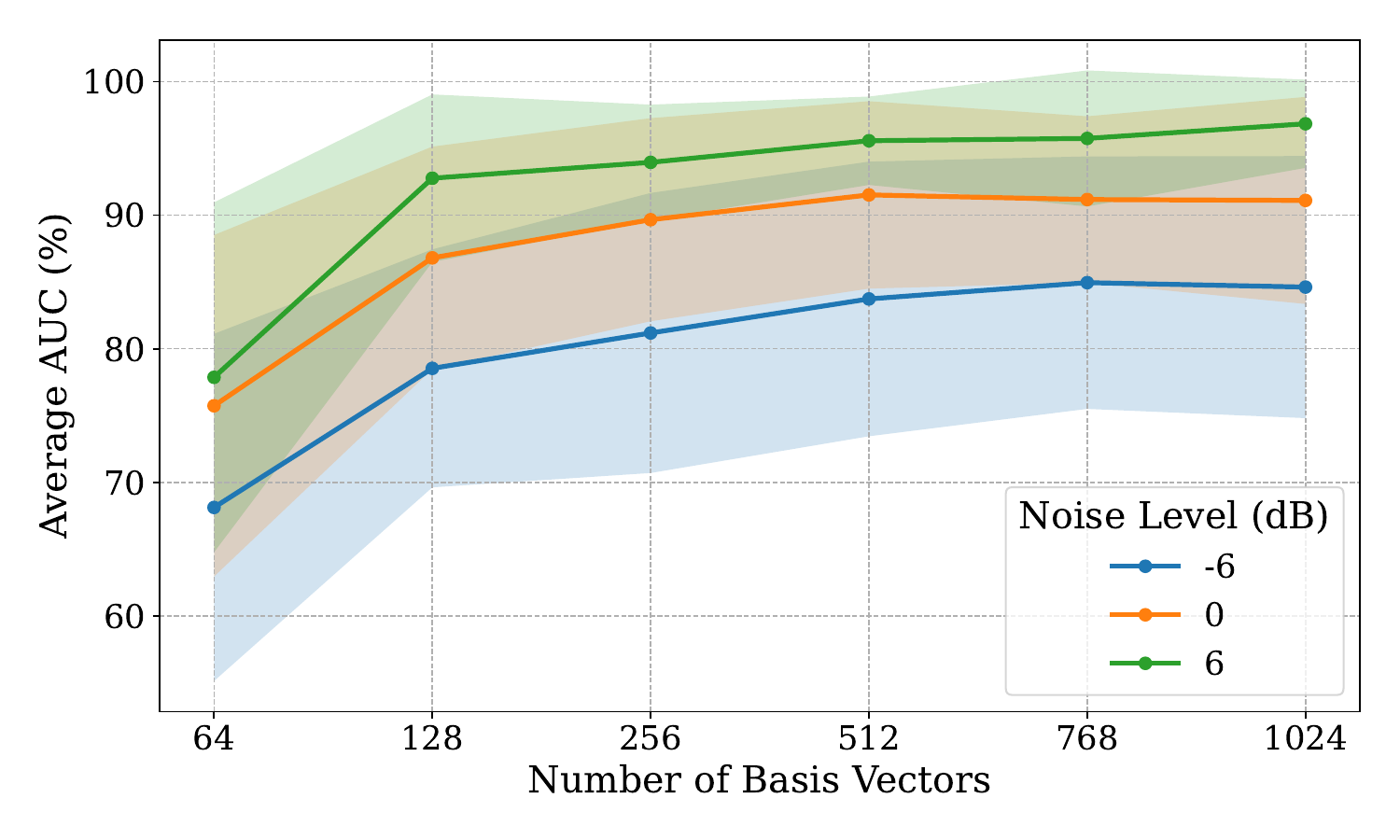}
            \caption{Effect of memory sizes.}\label{fig:nmf_order}
        \end{subfigure}
        &
        \begin{subfigure}[t]{0.325\textwidth}
            \centering
            \includegraphics[width=\linewidth,height=5cm,keepaspectratio]{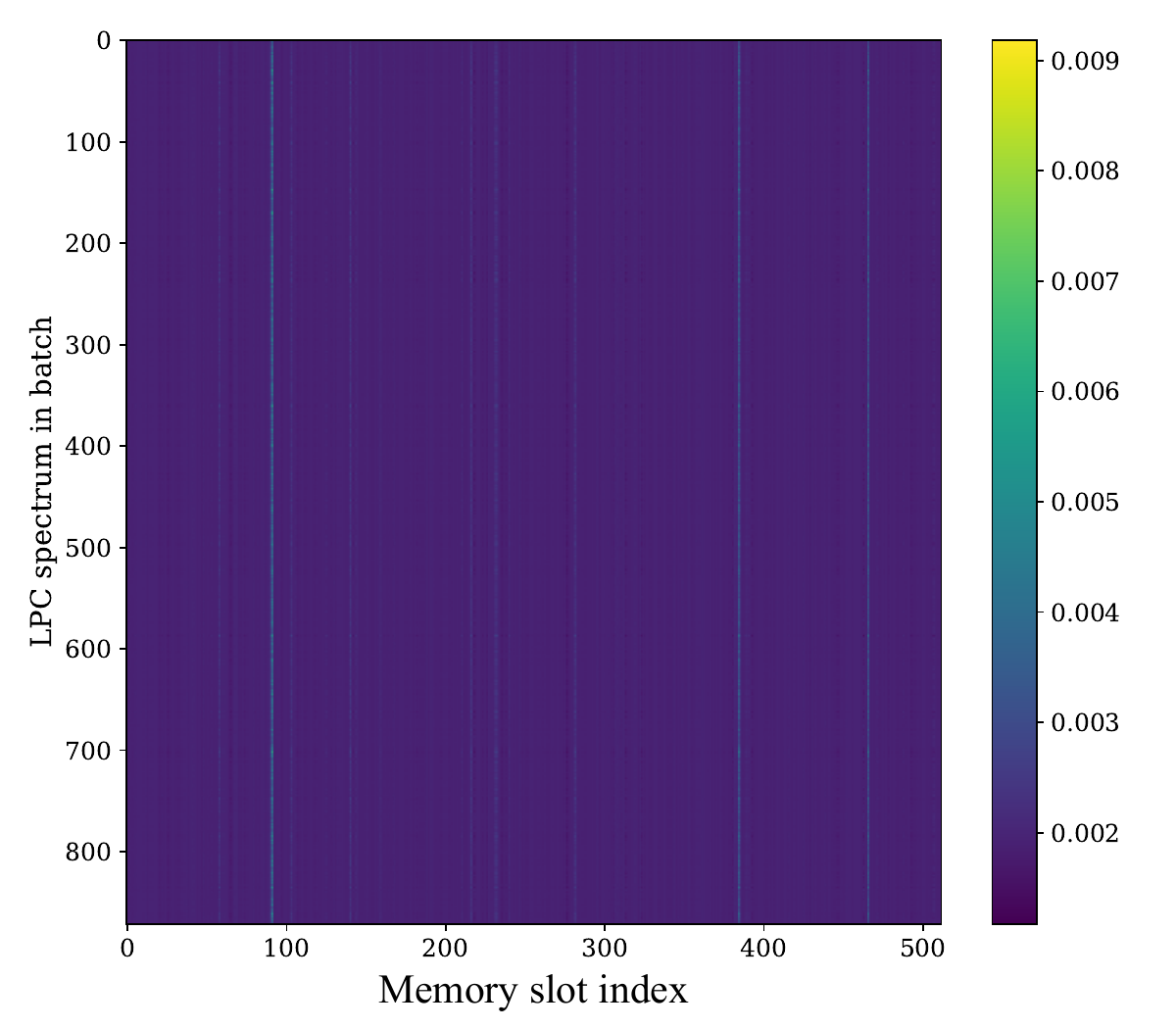}
            \caption{Memory-item activation heatmap without BatchNorm (Valve sec.~06, $-6$ dB, MIMII).}\label{fig:wo_BN}
        \end{subfigure}
        &
        \begin{subfigure}[t]{0.325\textwidth}
            \centering
            \includegraphics[width=\linewidth,height=5cm,keepaspectratio]{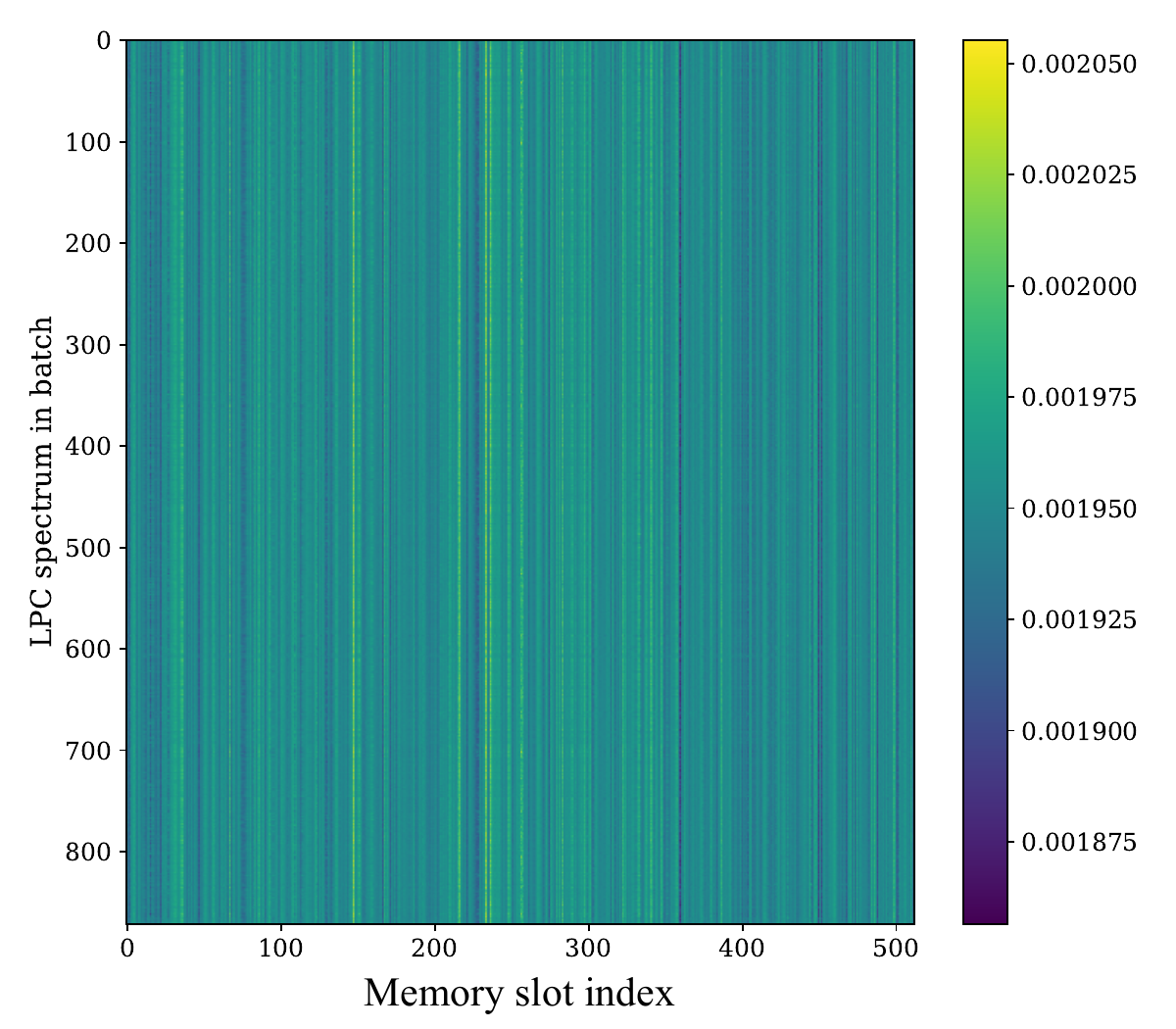}
            \caption{Memory-item activation heatmap with BatchNorm (Valve sec.~06, $-6$ dB, MIMII).}\label{fig:w_BN}
        \end{subfigure}
    \end{tabular}

    \caption{(a) Effect of memory size and memory-item activation heatmaps (b) without BatchNorm vs.\ (c) with BatchNorm.}
    \label{fig:compare_bn}
    \vspace{-3mm}
\end{figure*}

\subsection{Main results on DCASE2020T2}
We first isolated the impact of the input representation. Table \ref{tab:dcase2020_mimii_subset} compares a standard AE trained with spectrogram features (Log-Mel) against the same AE trained with LPC-spectrum features on \texttt{sec-dev}. Replacing spectrograms with the LPC spectrum yielded large gains on \emph{non-stationary} machine types such as Slider and Valve, suggesting that emphasizing the spectral envelope can be more robust than relying on highly time-varying spectrogram patterns. For Toy-car and Fan, however, the LPC spectrum did not uniformly improve the AE baseline, whereas MemNMF improved performance, suggesting that the training setup (e.g., hyperparameters) may be sub-optimal for these cases. In addition, Toy-conveyor degraded for both AE and MemNMF relative to Log-Mel, highlighting a potential limitation of LPC features on this machine type. Prior work also notes Toy-conveyor as a difficult case for learned ASD representations~\cite{wilkinghoff_subcluster_2021}. Compared with Log-Mel spectrograms, the LPC spectrum smooths time--frequency detail into a global envelope; if Toy-conveyor anomalies are localized or overlap spectrally with normal machine noise, this smoothing can attenuate discriminative cues and reduce normal--anomaly separation.

Table \ref{tab:results_all} summarizes the overall benchmark results on DCASE2020T2 across \texttt{sec-dev} and \texttt{sec-eval} using the arithmetic mean of AUC and pAUC (Amean). On this aggregate metric, our MemNMF achieved consistent improvements over the AE baseline and competitive performance against recent AE-based approaches (e.g., ANP-IDNN). On \texttt{sec-dev}, MemNMF was competitive with PAE, which uses a stronger Transformer-based architecture, suggesting that a substantial portion of the gain can be attributed to the LPC-spectrum input and the proposed memory-based reconstruction. However, MemNMF underperformed AudDSR, which can benefit from training on broader machine data and from anomaly augmentation strategies, whereas our setting focuses on learning only from the target normal data within each machine section.

\begin{table}[!t]
    \centering
    \caption{AUC scores for each machine type on the MIMII \cite{purohit_mimii_2019} dataset.}
    \label{tab:main_result}
    \scriptsize
    \resizebox{\columnwidth}{!}{%
        \begin{tabular}{lcccccc}
            \toprule
            Model & AE~\cite{regional_local} & VAE~\cite{regional_local} & GRLNet~\cite{regional_local} & MemNMF-512 & MemNMF-768 \\
            \midrule
            \multicolumn{6}{c}{\textbf{-6 dB}} \\
            \midrule
            Fan & 68.7 & 71.5 & 69.9 & 74.2 & \textbf{76.4} \\
            Pump & 71.0 & 71.0 & \textbf{77.5} & 75.7 & 77.3 \\
            Slider & 73.4 & 70.5 & 75.0 & 93.6 & \textbf{94.1} \\
            Valve & 50.3 & 49.8 & 53.4 & 91.4 & \textbf{92.0} \\
            Avg. & 65.9 & 65.7 & 68.9 & 83.7 & \textbf{85.0} \\
            \midrule
            \multicolumn{6}{c}{\textbf{0 dB}} \\
            \midrule
            Fan & 84.9 & 84.9 & 86.6 & 86.1 & \textbf{87.3} \\
            Pump & 81.6 & 81.7 & 85.3 & \textbf{86.8} & 83.5 \\
            Slider & 78.5 & 78.1 & 80.9 & 97.0 & \textbf{97.2} \\
            Valve & 54.9 & 54.7 & 57.0 & 96.2 & \textbf{96.7} \\
            Avg. & 75.0 & 74.8 & 77.5 & \textbf{91.5} & 91.2 \\
            \midrule
            \multicolumn{6}{c}{\textbf{6 dB}} \\
            \midrule
            Fan & 95.3 & 94.8 & \textbf{95.3} & 93.5 & 92.9 \\
            Pump & 86.9 & 87.7 & 90.1 & 92.2 & \textbf{93.0} \\
            Slider & 90.3 & 89.7 & 91.1 & 98.3 & \textbf{98.7} \\
            Valve & 59.5 & 57.3 & 63.9 & \textbf{98.4} & \textbf{98.4} \\
            Avg. & 83.0 & 82.4 & 85.1 & 95.6 & \textbf{95.7} \\
            \midrule
            Avg. (all) & 74.6 & 74.3 & 77.2 & 90.4 & \textbf{90.6} \\
            \bottomrule
        \end{tabular}
    }
    \vspace{-2mm}
\end{table}

\subsection{Results on MIMII}
The performance of our method under different SNRs on the MIMII dataset is shown in Table~\ref{tab:main_result}. Using 768 memory components yielded the best overall result, achieving 90.6\% average AUC across machine types and SNRs, and outperforming GRLNet \cite{regional_local} (another AE-based approach) by 13.4\% (90.6\% vs.\ 77.2\%). Notably, on the non-stationary machine types (Valve and Slider), MemNMF maintained over 90\% AUC even at -6~dB, indicating strong robustness in noisy conditions.

\subsubsection{Memory sizes}
We varied the memory size (equal to the number of NMF components) and found that performance increased with more components, peaking around 512--768; larger sizes brought little additional gain (Fig.~\ref{fig:nmf_order}).

\subsubsection{Design choices}
Table~\ref{tab:design_choices} compares the default MemNMF with trained memory-network ablations and NMF-only baselines. In the default model, NMF first learns \(B\) and \(B_p=\operatorname{pinv}(B)\), which are kept fixed while the memory network uses \(K=B_pP_K\), \(V=B^\top P_V\), and BatchNorm on the logits \(\mathbf{X}K\) before Softmax; only \(P_K\), \(P_V\), and BatchNorm parameters were trained. The ``use \(B^T P_K\) as \(K\)'' row replaces only the default key \(B_pP_K\), while ``w/o BatchNorm'' removes only this logit normalization. The ``w/o NMF'' row keeps the same memory-network training but randomly initializes the key/value dictionaries instead of anchoring them to NMF-derived \(B_p\) and \(B\). The last two rows are NMF-only baselines without memory-network fine-tuning: NNLS estimates test activations by solving a non-negative reconstruction problem with fixed \(B\), while \(B_p\) gives an unconstrained pseudo-inverse encoding. NNLS substantially outperforms \(B_p\)-only reconstruction, confirming a strong fixed NMF baseline, but fits each test sample independently. MemNMF performs better, especially at low SNRs, by learning a BatchNorm-normalized memory lookup for more robust anomaly-score separation. Fig.~\ref{fig:w_BN} shows that BatchNorm decreases activation sparsity, allowing the memory network to utilize a wider variety of memory items for reconstruction, consistent with prior findings~\cite{Bjorck2018UnderstandingBN}. Finally, using \(B_p\) to initialize the Key (instead of \(B^T P_K\)) consistently yields better results, implying that the pseudo-inverse basis provides a stronger encoding for computing activations.

\begin{table}[!t]
    \centering
    \scriptsize
    \caption{AUC scores for MemNMF ablations and NMF-only baselines.}\label{tab:design_choices}
    \scalebox{1}{
    \begin{tabular}{@{}llll@{}}
        \toprule
        Method & -6 dB & 0 dB & 6 dB \\
        \midrule
        MemNMF & 84.1 & 91.5 & 95.6 \\
        \hspace{1em}\textit{use \(B^T P_K\) as \(K\)} & 82.3 \ (\textcolor{red}{-1.8}) & 88.7 \ (\textcolor{red}{-2.8}) & 93.3 \ (\textcolor{red}{-2.3}) \\
        \hspace{1em}\textit{w/o BatchNorm} & 77.6 \ (\textcolor{red}{-6.5}) & 87.0 \ (\textcolor{red}{-4.5}) & 94.7 \ (\textcolor{red}{-0.9}) \\
        \hspace{1em}\textit{w/o NMF (Random Init.)} & 67.9 \ (\textcolor{red}{-16.2}) & 79.2 \ (\textcolor{red}{-12.3}) & 83.3 \ (\textcolor{red}{-12.3}) \\
        NMF w/ \(B_p\) & 61.8 \ (\textcolor{red}{-22.3}) & 63.1 \ (\textcolor{red}{-28.4}) & 64.3 \ (\textcolor{red}{-31.3}) \\
        NMF w/ NNLS & 78.5 \ (\textcolor{red}{-5.6}) & 87.5 \ (\textcolor{red}{-4.0}) & 94.6 \ (\textcolor{red}{-1.0}) \\
        \bottomrule
    \end{tabular}
    }
    \vspace{-4mm}
\end{table}


\section{Conclusion}
We presented MemNMF, aiming to address limitations of conventional autoencoder frameworks for ASD, by reconstructing LPC spectra using a memory module initialized from an NMF dictionary, thereby constraining reconstruction to sparse, non-negative combinations of normal bases. Experiments on MIMII and DCASE2020 Task~2 show consistent gains across machine types and operating conditions, with strong robustness in noisy, non-stationary settings. Future work will study better transfer across domains and stronger constraints for reliability under distribution shift.

\section*{Acknowledgment}
This work was supported by JTEKT.

\FloatBarrier
\bibliographystyle{IEEEtran}
\bstctlcite{BSTcontrol}
\bibliography{references}

@IEEEtranBSTCTL{IEEEexample:BSTcontrol,
  CTLuse_forced_etal       = "yes",
  CTLmax_names_forced_etal = "5",
  CTLnames_show_etal       = "2" ,
  CTLuse_url = "no",
  CTLuse_month = "no",
  CTLuse_editor = "no",
CTLuse_publisher = "no",
}

@inproceedings{huang_network_2006,
  author    = {Huang, Ling and Nguyen, XuanLong and Garofalakis, Minos N. and Jordan, Michael I. and Joseph, Anthony D. and Taft, Nina},
  title     = {In-Network PCA and Anomaly Detection},
  booktitle = {Advances in Neural Information Processing Systems},
  volume    = {19},
  pages     = {617--624},
  year      = {2006}
}

@book{mallat2009wavelet,
  title={A Wavelet Tour of Signal Processing: The Sparse Way},
  author={Mallat, Stephane},
  year={2009},
  publisher={Academic Press},
  edition={3rd},
  isbn={978-0-12-374370-1}
}

@article{lee1999nmf,
  author  = {Lee, Daniel D. and Seung, H. Sebastian},
  title   = {Learning the Parts of Objects by Non-negative Matrix Factorization},
  journal = {Nature},
  volume  = {401},
  number  = {6755},
  pages   = {788--791},
  year    = {1999},
  doi     = {10.1038/44565}
}

@inproceedings{lee2000algorithms,
  author    = {Lee, Daniel D. and Seung, H. Sebastian},
  title     = {Algorithms for Non-negative Matrix Factorization},
  booktitle = {Advances in Neural Information Processing Systems},
  volume    = {13},
  year      = {2000}
}

@inproceedings{Purohit_DCASE2019_01,
    Author = "Purohit, Harsh and Tanabe, Ryo and Ichige, Takeshi and Endo, Takashi and Nikaido, Yuki and Suefusa, Kaori and Kawaguchi, Yohei",
    title = "{MIMII Dataset}: Sound Dataset for Malfunctioning Industrial Machine Investigation and Inspection",
    year = "2019",
    booktitle = "Proc. DCASE2019",
    month = "November",
    pages = "209--213",
    url = "http://dcase.community/documents/workshop2019/proceedings/DCASE2019Workshop\_Purohit\_21.pdf"
}

@inproceedings{gong_memorizing_2019,
	address = {Seoul, Korea (South)},
	title = {Memorizing {Normality} to {Detect} {Anomaly}: {Memory}-{Augmented} {Deep} {Autoencoder} for {Unsupervised} {Anomaly} {Detection}},
	copyright = {https://ieeexplore.ieee.org/Xplorehelp/downloads/license-information/IEEE.html},
	isbn = {978-1-72814-803-8},
	shorttitle = {Memorizing {Normality} to {Detect} {Anomaly}},
	url = {https://ieeexplore.ieee.org/document/9010977/},
	doi = {10.1109/ICCV.2019.00179},
	language = {en},
	urldate = {2024-05-09},
	booktitle = {Proc. ICCV},
	publisher = {IEEE},
	author = {Gong, Dong and Liu, Lingqiao and Le, Vuong and Saha, Budhaditya and Mansour, Moussa Reda and Venkatesh, Svetha and Van Den Hengel, Anton},
	month = oct,
	year = {2019},
	pages = {1705--1714},
}

@inproceedings{wilkinghoff_subcluster_2021,
	title = {Sub-{Cluster} {AdaCos}: {Learning} {Representations} for {Anomalous} {Sound} {Detection}},
	isbn = {978-1-66543-900-8},
	shorttitle = {Sub-{Cluster} {AdaCos}},
	url = {https://ieeexplore.ieee.org/document/9534290/},
	doi = {10.1109/IJCNN52387.2021.9534290},
	language = {en},
	urldate = {2024-01-12},
	booktitle = {Proc. IJCNN},
	author = {Wilkinghoff, Kevin},
	year = {2021},
	pages = {1--8},
}

@inproceedings{wichern_anomalous_2021,
	title = {Anomalous {Sound} {Detection} {Using} {Attentive} {Neural} {Processes}},
	url = {https://ieeexplore.ieee.org/document/9632762/},
	doi = {10.1109/WASPAA52581.2021.9632762},
	language = {en},
	urldate = {2024-03-12},
	booktitle = {Proc. WASPAA},
	author = {Wichern, Gordon and Chakrabarty, Ankush and Wang, Zhong-Qiu and Roux, Jonathan Le},

	year = {2021},
	pages = {186--190},
}

@inproceedings{suefusa_anomalous_2020,
	title = {Anomalous {Sound} {Detection} {Based} on {Interpolation} {Deep} {Neural} {Network}},
	url = {https://ieeexplore.ieee.org/abstract/document/9054344},
	doi = {10.1109/ICASSP40776.2020.9054344},
	urldate = {2024-01-11},
	booktitle = {Proc. ICASSP},
	author = {Suefusa, Kaori and Nishida, Tomoya and Purohit, Harsh and Tanabe, Ryo and Endo, Takashi and Kawaguchi, Yohei},

	year = {2020},
	pages = {271--275},
}

@inproceedings{koizumi_description_2020,
	title = {Description and {Discussion} on {DCASE2020} {Challenge} {Task2}: {Unsupervised} {Anomalous} {Sound} {Detection} for {Machine} {Condition} {Monitoring}},
	shorttitle = {Description and {Discussion} on {DCASE2020} {Challenge} {Task2}},
	url = {http://arxiv.org/abs/2006.05822},
	urldate = {2023-11-04},
	publisher = {arXiv},
	author = {Koizumi, Yuma and Kawaguchi, Yohei and Imoto, Keisuke and Nakamura, Toshiki and Nikaido, Yuki and Tanabe, Ryo and Purohit, Harsh and Suefusa, Kaori and Endo, Takashi and Yasuda, Masahiro and Harada, Noboru},
	year = {2020},
	booktitle= {Proc. DCASE},
}

@inproceedings{koizumi_toyadmos_2019,
    Author = "Koizumi, Yuma and Saito, Shoichiro and Uematsu, Hisashi and Harada, Noboru and Imoto, Keisuke",
    title = "{ToyADMOS}: A Dataset of Miniature-machine Operating Sounds for Anomalous Sound Detection",
    year = "2019",
    booktitle = "Proc. WASPAA",

    pages = "308--312",
    url = "https://ieeexplore.ieee.org/document/8937164"
}

@inproceedings{purohit_mimii_2019,
    Author = "Purohit, Harsh and Tanabe, Ryo and Ichige, Takeshi and Endo, Takashi and Nikaido, Yuki and Suefusa, Kaori and Kawaguchi, Yohei",
    title = "{MIMII Dataset}: Sound Dataset for Malfunctioning Industrial Machine Investigation and Inspection",
    year = "2019",
    booktitle = "Proc. DCASE",

    pages = "209--213",
}

@IEEEtranBSTCTL{BSTcontrol,
  CTLuse_forced_etal       = "yes",
  CTLmax_names_forced_etal = "3",
  CTLnames_show_etal       = "1"
}

@article{Harada_EUSIPCO2023_01,
    author = "Harada, Noboru and Niizumi, Daisuke and Takeuchi, Daiki and Ohishi, Yasunori and Yasuda, Masahiro",
    title = "First-Shot Anomaly Detection for Machine Condition Monitoring: A Domain Generalization Baseline",
    journal = "Proc. EUSIPCO",
    pages = "191--195",
    year = "2023"
}

@techreport{GleichmannTNT2024,
    Author = "Gleichmann, Lars Christian and Adhisantoso, Yeremia Gunawan and Lange, Alexander and Le Xuan, Quy",
    title = "Anomaly Sound Detector Based on Variational Autoencoder with Hyperparameter Optimization Strategy",
    institution = "DCASE2024 Challenge",
    year = "2024"
}

@INPROCEEDINGS{local_den_norm_kevin,
  author={Wilkinghoff, Kevin and Yang, Haici and Ebbers, Janek and Germain, François G. and Wichern, Gordon and Roux, Jonathan Le},
  booktitle={Proc. ICASSP}, 
  title={Keeping the Balance: Anomaly Score Calculation for Domain Generalization}, 
  year={2025},
  volume={},
  number={},
  pages={1-5},
  doi={10.1109/ICASSP49660.2025.10888402}}

@inproceedings{Bjorck2018UnderstandingBN,
  title={Understanding Batch Normalization},
  author={Johan Bjorck and Carla Pedro Gomes and Bart Selman},
  booktitle={Proc. NIPS},
  year={2018},
  url={https://api.semanticscholar.org/CorpusID:46956723}
}

@Inproceedings{koizumi_spidernet_9053620,
  author={Koizumi, Yuma and Yasuda, Masahiro and Murata, Shin and Saito, Shoichiro and Uematsu, Hisashi and Harada, Noboru},
  booktitle={Proc. ICASSP}, 
  title={SPIDERnet: Attention Network For One-Shot Anomaly Detection In Sounds}, 
  year={2020},
  volume={},
  number={},
  pages={281-285},
  doi={10.1109/ICASSP40776.2020.9053620}}

@article{saito1967theoretical,
  author    = {S. Saito and F. Itakura},
  title     = {Theoretical consideration of the statistical optimum recognition of the spectral density of speech},
  journal   = {J. Acoust. Soc. Japan},
  year      = {1967},
  month     = {January}
}

@ARTICLE{lpc_review_1451722,
  author={Makhoul, J.},
  journal={Proceedings of the IEEE}, 
  title={Linear prediction: A tutorial review}, 
  year={1975},
  volume={63},
  number={4},
  pages={561-580},
  doi={10.1109/PROC.1975.9792}}

@INPROCEEDINGS{Koizumi_8081297,
  author={Koizumi, Yuma and Saito, Shoichiro and Uematsu, Hisashi and Harada, Noboru},
  booktitle={Proc. EUSIPCO}, 
  title={Optimizing acoustic feature extractor for anomalous sound detection based on Neyman-Pearson lemma}, 
  year={2017},
  volume={},
  number={},
  pages={698-702},
  doi={10.23919/EUSIPCO.2017.8081297}}

@article{Uematsu2017AnomalyDT,
  title={Anomaly Detection Technique in Sound to Detect Faulty Equipment},
  author={Hisashi Uematsu and Yuma Koizumi and Shoichiro Saito and Akira Nakagawa and Noboru Harada},
  journal={NTT Technical Review},
  year={2017},
  url={https://api.semanticscholar.org/CorpusID:195293628}
}

@article{andersen1974calculation,
  title={On the calculation of filter coefficients for maximum entropy spectral analysis},
  author={Andersen, N},
  journal={Geophysics},
  volume={39},
  number={1},
  pages={69--72},
  year={1974},
  publisher={Society of Exploration Geophysicists}
}

@article{burg1975maximum,
  title={Maximum entropy spectral analysis.},
  author={Burg, John Parker},
  journal={Ph. D. thesis, Stanford University, Department of Geophysics},
  year={1975},
  publisher={Ph. D. thesis, Stanford University, Department of Geophysics}
}

@InProceedings{ brian_mcfee-proc-scipy-2015,
  author    = { {B}rian {M}c{F}ee and {C}olin {R}affel and {D}awen {L}iang and {D}aniel {P}.{W}. {E}llis and {M}att {M}c{V}icar and {E}ric {B}attenberg and {O}riol {N}ieto },
  title     = { librosa: {A}udio and {M}usic {S}ignal {A}nalysis in {P}ython },
    booktitle = {Proc. SciPy},
  pages     = { 18 - 24 },
  year      = { 2015 },
  doi       = { 10.25080/Majora-7b98e3ed-003 }
}

@inproceedings{
    loshchilov2017sgdr,
    title={{SGDR}: Stochastic Gradient Descent with Warm Restarts},
    author={Ilya Loshchilov and Frank Hutter},
    booktitle = {Proc. ICLR},
    year={2017},
    url={https://openreview.net/forum?id=Skq89Scxx}
}

@article{Kingma2014AdamAM,
  title={Adam: A Method for Stochastic Optimization},
  author={Diederik P. Kingma and Jimmy Ba},
  journal={CoRR},
  year={2014},
  volume={abs/1412.6980},
  url={https://api.semanticscholar.org/CorpusID:6628106}
}

@article{Cichocki2009FastLA,
  title={Fast Local Algorithms for Large Scale Nonnegative Matrix and Tensor Factorizations},
  author={Andrzej Cichocki and A. Phan},
  journal={IEICE Trans. Fundam. Electron. Commun. Comput. Sci.},
  year={2009},
  volume={92-A},
  pages={708-721},
  url={https://api.semanticscholar.org/CorpusID:16858053}
}

@inproceedings{regional_local,
author = {Sha, Yu and Gou, Shuiping and Faber, Johannes and Liu, Bo and Li, Wei and Schramm, Stefan and Stoecker, Horst and Steckenreiter, Thomas and Vnucec, Domagoj and Wetzstein, Nadine and Widl, Andreas and Zhou, Kai},
title = {Regional-Local Adversarially Learned One-Class Classifier Anomalous Sound Detection in Global Long-Term Space},
year = {2022},
url = {https://doi.org/10.1145/3534678.3539133},
doi = {10.1145/3534678.3539133},
booktitle = {Proc. ACM SIGKDD},
pages = {3858–3868},
numpages = {11},
}

@INPROCEEDINGS{zeng_joint_2023,
  author={Zeng, Xiao-Min and Song, Yan and Zhuo, Zhu and Zhou, Yu and Li, Yu-Hong and Xue, Hui and Dai, Li-Rong and McLoughlin, Ian},
  booktitle={Proc. ICASSP}, 
  title={Joint Generative-Contrastive Representation Learning for Anomalous Sound Detection}, 
  year={2023},
  volume={},
  number={},
  pages={1-5},
  doi={10.1109/ICASSP49357.2023.10095568}}

@INPROCEEDINGS{zavrrtanki_10447941,
  author={Zavrtanik, Vitjan and Marolt, Matija and Kristan, Matej and Skočaj, Danijel},
  booktitle={Proc. ICASSP}, 
  title={Anomalous Sound Detection by Feature-Level Anomaly Simulation}, 
  year={2024},
  volume={},
  number={},
  pages={1466-1470},
  doi={10.1109/ICASSP48485.2024.10447941}}

\end{document}